# Direct Measurement of the Surface Tension of a Soft Elastic Hydrogel: Exploration of Elasto-Capillary Instability in Adhesion

Aditi Chakrabarti and Manoj K. Chaudhury[*]

Department of Chemical Engineering
Lehigh University, Bethlehem, PA 18015

**ABSTRACT:** An adhesively stressed thin film of a soft hydrogel confined between two rigid flat substrates auto-roughens with its dominant wavelength ($\lambda$) exhibiting pronounced dependence on the film thickness ($H$). A linear stability analysis confirmed that this long wavelength instability ($\lambda \sim 7H$) is due to an elasto-capillary effect, the implementation of which required direct measurements of the surface tension and the elasticity of the gel. The surface tension of the gel was estimated from the fundamental spherical harmonic of a hemispherical cap of the gel that was excited by an external noise. The shear modulus ($\mu$) of the gel was determined from its resonant shear mode in a confined geometry. During the course of this study, it was found that a high density steel ball submerges itself inside the gel by balancing its excess weight with the accumulated strain induced elastic force that allows another estimation of its elastic modulus. The large ratio (1.8 mm) of the surface tension to its elasticity ascertains the role of elasto-capillarity in the adhesion induced pattern formation with such gels. Experimental results are in accord with a linear stability analysis that predicts that the rescaled wavelength $\lambda(\mu H/\gamma)^{0.27}$ is linear with $H$, which modifies the conventional stress to pull-off a rigid flat object from a very soft film by a multiplicative factor: $(\gamma/\mu H)^{1/4}$. The analysis also suggests some new results related to the role of the finite dilation of a material in interfacial pattern formation that may have non-trivial consequences in the adhesive delamination of very thin and/or soft elastic films via self-generated cracks.

* e-mail: mkc4@lehigh.edu

## 1. INTRODUCTION

The joint roles of the surface tension and the elastic forces have long been recognized in soft matter physics. Starting with the original proposal of Lester[1] that a soft solid can be deformed by the normal component of the surface tension of a liquid drop, the subject has continued to blossom[2-8] till date with the identification of a scale of the surface deformation in terms of the surface tension ($\gamma$) divided by the elasticity ($\mu$). This so called elasto-capillary length ($\gamma/\mu$) also appears in various other surface phenomena such as the wrapping of a liquid drop by a thin elastic film[9], coalescence of thin wet fibers[10,11], buckling of thin rods inside a liquid drop[12], cavitation in soft hydrogel[13,14], bulging[15] of a thin elastic channel due to capillary pressure, flattening[16] of a soft solid by surface tension, and Rayleigh instability[17] in a soft gel to name a few. Elasto-capillary effect has also been found to be important in the nucleation of creases[18] in soft solid and it manifests directly in the form of the radius ($a$) of the contact deformation[19] of two spheres with an effective radius $R$ as $a^3/R^2 \sim \gamma/\mu$, which can be used as a sensitive probe to study surfactant adsorption[20] related capillary effects as well.

The objective of this paper is to show that elasto-capillary instability can play an important role in the formation of self-generated cracks in soft confined films. Usually, when a rigid substrate adhered to a thin rubber film is subjected to a tensile stress, the interface ceases to be flat[21-27]. As was discovered (or observed) independently at the Lehigh University[21] and at the University of Ulm[22], the entire surface or the line of contact of such an adhesively stressed film roughens with a characteristic length scale that is simply proportional to the thickness of the film (*H*). This happens when the elasto-capillary number ($\mu H/\gamma$) of the film is much larger than unity. On the other hand, if $\gamma/\mu$ is comparable to the film thickness, a material length scale emerges that moderates the interfacial roughness. A simple scaling analysis[26] suggests (see Appendix A) that the wavelength of the instability would be of the following form: $\lambda \sim H (1+ \gamma/\mu H)^{1/4}$, according to which there are three distinct regimes of elastic instability. For a high modulus film ($\gamma/\mu \ll H$), the wavelength is proportional to thickness, i.e. $\lambda \sim H$. When $\gamma/\mu$ is slightly smaller than *H*, the instability wavelength depends on both the geometric and the material scales ($\gamma/\mu$) almost additively: $\lambda \sim H + \gamma/4\mu$. On the other hand, when $\gamma/\mu \gg H$, the elasto-capillary and the geometric scales are strongly coupled as $\lambda \sim (\gamma H^3/\mu)^{1/4}$. Gonuguntla et al[28] studied this regime using ultra-thin films of elastomeric PDMS, in which the patterns were frozen by UV ozone treatment and analyzed after the contactor was removed. They showed that the wavelength of instability deviates from the conventional $\lambda \sim 3H$ relationship. While this was the first attempt to document the elasto-capillarity in adhesion induced pattern formation, there are certain concerns with the way these experiments were performed and interpreted. To begin with, since an UV-ozone treatment could modify the properties of the film to a substantial depth in a gradient fashion, it is not clear *a priori* whether the modulus of the ultra-thin film would be same as that of the untreated bulk elastomer. In addition, the surface tension of the ultra-thin elastomeric PDMS was assumed to be same as that in its liquid state, which cannot be guaranteed, either in the native state and, especially, when its surface is post-hardened. Most importantly, however, as the measurements were performed *ex-situ* after preserving the pattern and then removing the contactor, it is not at all clear whether the long wave features of the instability were the reminiscence of the surface tension induced flattening[16], in which the short wave features decay, or it was indeed due to the adhesion induced instability [28]. Whilst these criticisms do not take away the novelty and the elegance of these experiments[28], there is, nevertheless, a need to conduct definitive experiments and to carry out the related analysis on such types of instabilities in a system where the patterns can be observed *in situ* and where the solid surface tension and the elasticity of the deformable adhesive can be measured independently.

The purpose of this paper is to report such measurements performed with a physically cross-linked polyacrylamide hydrogel[29-31], the elastic modulus of which could easily be controlled and set to a rather low value. A further inspiration for such a study stems from the fact that these types of ultra-soft gels are increasingly used in various biomedical, cosmetic and adhesive technologies[32]. As the deformability is a major issue in these studies, the soft gels have become the testing grounds for various types of mechanical and rheological characterizations[33] over the years. In spite of considerable progress, however, characterizations of the ultra-soft gels can be quite challenging in certain settings, especially when the elastic forces are comparable to that of capillarity. The interplay of these factors, nevertheless, makes these gels interesting candidates of study in an evolving branch of rheology where surface tension, elasticity and viscosity play their respective roles.

In order to observe the putative elasto-capillary instability and interpret it on a sound physical ground, we had to accomplish three different but related objectives. The main objective was to design an experiment with which the instability could be induced and measured *in situ* over a considerable range of film thickness. This objective was accomplished with a gel confined in a wedge shaped geometry, the thickness of which varied from about 0 micron to about 180 micron in a linear fashion. This idea of using a thickness gradient is philosophically similar to that of Stafford et al[34] in studying the effect of thickness of the top layer in wrinkling instability.

A linear stability analysis was performed to understand the long wavelength feature of the elasto-capillary instability, the execution of which, however, required values of the surface tension and the elasticity of the physically cross-linked hydrogel. Its surface tension was estimated from the spherical harmonic of the free surface of a hemispherical gel cap, whereas its elasticity was estimated from the natural shear resonance mode of a hydrogel slab after submitting each to a random mechanical vibration. In connection with measuring the elasticity of the gel, we also report a novel observation, in which a steel ball remains suspended in the gel by balancing its weight with the accumulated shear strain induced elastic force of the surrounding medium. This is the static or a self-braking version of the classical Stokes experiment that allowed estimation of the shear modulus of the gel both at small as well as at large deformations in the absence of dynamics. Once the surface tension and the elasticity of the gel were measured, an elasto-capillary length could be estimated and compared with that obtained from the adhesion induced instability patterns.

The paper is organized as follows. After describing the experimental protocols, we discuss the methods to measure the elasticity and the surface tension of an ultra-soft hydrogel. These measurements established an elasto-capillary length of the hydrogel, which was then used to interpret the spatial wavelength of the adhesion induced interfacial instability. The paper is then concluded with a discussion on how elasto-capillarity could be an important factor in deciding the failure modes of the interface of an elastic film sandwiched between two rigid substrates.

## 2. EXPERIMENTAL SECTION

### 2.1. MATERIALS

The chemicals used for the preparation of the hydrogel were N-(hydroxymethyl)-acrylamide (48% solution in water, Sigma Aldrich®), potassium persulphate (99.99% trace metals basis, Sigma Aldrich®) and N,N,N',N'-Tetramethylethylenediamine (TEMED, ≥99.5%, purified by re-distillation, Sigma Aldrich®). Deionized water (DI water) was obtained from Thermo Scientific® Barnstead E-pure* unit. Glass slides of two different sizes (75 mm x 50 mm x 1 mm) and (75 mm x 25 mm x 1 mm) were obtained from Fisher Scientific (Fisherbrand® Microscopic slides). The glass cover slips were purchased from Corning (Corning Cover Glass, No. 1, 24 mm x 60 mm). Freshly opened Fisherbrand® Borosilicate Glass vials (27 mm diameter x 70 mm high) were already quite clean; but they were further washed thoroughly with DI (deionized) water and blow dried with ultra-high purity Nitrogen gas. The gel solutions were prepared in these vials. Steel balls (Bearing-Quality E52100 Alloy Steel, Hardened Ball) of diameters ranging from 1 mm to 10 mm were purchased from McMaster-Carr®. These balls were sonicated in Acetone (General use HPLC-UV grade, Pharmco Aaper®) in a Fisher Scientific Ultrasonic Cleaner (Model no. FS5) for 10 minutes after which they were blow dried with pure Nitrogen gas. In some of the experiments, glass slides were reacted with a silane [dodecyltrichlorosilane (HC-12, Gelest Inc), hexadecyltrichlorosilane (HC-16, Gelest Inc.) or *1H,1H,2H,2H*-perfluorodecyltrichlorosilane (FC-10, Alfa Aesar)], the details of which were reported previously[35]. For some of the experiments, glass slides were modified with a thin (5 nm) film of polydimethylsiloxane (PDMS, Gelest DMS-T22) using a method reported in the literature[36].

2.2. PREPARATION OF GEL

The physically cross-linked gel was prepared using a slight variation of the methods[29-31] reported in the literature in order to ensure that polymerization reaction could be carried out at room temperature in less than an hour. In a cleaned glass vial, N-(hydroxymethyl)-acrylamide and DI water were added to prepare a 3.5% (w/w) of the monomer in the solution, which was followed by degassing it with the bubbling of ultrapure nitrogen gas for 30 minutes while stirring it constantly with a magnetic stirrer. The polymerization reaction was initiated by first adding Potassium Persulphate (0.25 wt% basis) and then TEMED (0.3 wt% basis) to the above solution accompanied by constant stirring. The final solution was pipetted out of the vial and introduced in the respective setups soon after the TEMED was mixed. In all experiments, gelation reaction was carried out at room temperature for two hours, even though the reaction was complete within half an hour, which was ascertained from the fact that the modulus of the gel remains unchanged beyond this time.

## 2.3. MEASUREMENTS OF ELASTIC MODULI OF THE GEL

The shear modulus ($\mu$) was determined from the resonant shear mode of vibration of a gel slab confined between two parallel glass slides. One of the glass slides (75 mm x 25 mm x 1 mm) was coated with a monolayer of HC-16, which was then was placed above an untreated clean glass slide (75 mm x 50 mm x 1 mm) by maintaining a uniform gap of 1 mm between the two slides by means of two 1 mm thick spacers. The spacers themselves were prepared from the microscope glass slides, the edges of which were lined with thin strips of Teflon tapes for easy removal from the gel once the gelation was complete. This assembly was set up inside a polystyrene petri dish (VWR®, 150mm diameter, 15mm high) with stacks of deionized (DI) water soaked filter papers placed on the sides of the above assembly in order to create a humid environment (relative humidity of 99.9% at 23°C). The gel solution was inserted into the uniform gap between the slides by means of a sterile transfer pipette (7.7 mL, #202-1S, Thermo Scientific®, Samco*). One hydrophobic (above) and another hydrophilic (below) glass slides allowed the liquid to fill the gap by capillarity, but prevented its spreading beyond the edges of the Teflon coated spacers. This setup was left undisturbed for the next two hours while the gel slab (57 mm x 25 mm x 1 mm) cured, following which the spacers were removed. The lower plate of the assembly was fixed carefully onto the aluminum stage connected to a mechanical oscillator (Pasco Scientific, Model No: SF-9324) that was subjected to either a lateral or a vertical vibration using a Gaussian white noise (strength of 0.005 to 0.12 $m^2/s^3$). The noise was generated by a waveform generator (Agilent, model 33120A) and passed through an amplifier (Sherwood, Model No: RX-4105) before reaching the oscillator. The entire experimental setup was placed on a vibration isolation table (Micro-g, TMC). The shear and/or the vertical displacements of the upper glass slide were recorded with a high speed camera (Redlake Motion-Pro, Model no: 2000) operating at 1000 frames/s. The motion of the upper plate was later tracked using a MIDAS software (Midas2.0, Xcitex Inc., USA). The displacement fluctuations were fast Fourier transformed (FFT) using OriginLab® software to identify the resonant mode of vibration. The details of the basic methods can be found in previous publications[36,37].

## 2.4. STATIC STOKES' EXPERIMENT

In these experiments, the cleaned steel balls were gently dropped inside the glass vial containing the cured hydrogel. After submerging itself partially or fully inside the gel, the ball stood still inside the gel at a depth ($h$), which was captured by a Video Microscope (Infinity®) equipped with a CCD camera (jAi®, Model no. CV-S3200) with the help of WinTV application (Hauppauge®, USA) on the computer. Care was taken to ensure that the steel balls were at the centers of the vials to minimize wall effects. The images were analyzed using ImageJ® for calculating the depth of the steel ball into the gel. The calibration factor of the variable focal length microscope was determined from the known diameter of a steel ball in every run. Even though there was a minor distortion of the shape of the ball in the horizontal direction when viewed through the cylindrical glass vials, there was no such distortion in the vertical direction. All calibrations and measurements were carried out in vertical direction only.

2.5. DIRECT ESTIMATION OF THE SURFACE TENSION OF THE GEL USING VIBRATION

The surface tension was estimated from the vibration modes of the free surface of the hemispherical caps of the hydrogel prepared on hydrophobic glass slides. Glass slides were cut into small pieces (10 mm x 8 mm) using a diamond scriber which were then silanized by reacting them with the vapor of dodecyltrichlorosilane (HC-12). The pieces of these hydrophobic glass slides were fixed at the bases of small petri dishes (35 mm diameter x 10 mm high, Fisherbrand®) using a double sided Scotch® tape. After deposition of 2 to 40 $\mu$L size drops of the gel solution on these glass pieces (one drop per dish), the lids of the petri-dishes were closed. The filter papers placed on the sides of the petri-dishes were soaked with an aqueous solution of acrylamide monomer and TEMED with the same composition as the gel in order to suppress the evaporation of these ingredients from the gel drop itself.

The contact angle of the gel cap was ~90° on the silanized glass slide. The petri dishes were left undisturbed for 2 hours while the gel caps cured. After securely fixing the test substrate (the petridish with the samples inside it) on the aluminum stage of the mechanical oscillator, it was vibrated vertically with a Gaussian white noise (strength of 0.04 m$^2$/s$^3$). The height fluctuations of the gel caps were recorded with the high speed camera at 2000 frames/s which were subsequently analyzed with MIDAS 2.0. The fluctuations of the gel lenses were fast Fourier transformed (FFT) using OriginLab® software to identify the resonant mode of vibration.

2.6. ADHESION INSTABILITY EXPERIMENT

A hydrophobic glass slide (75 mm x 25 mm x 1 mm) was inclined above an untreated clean glass slide (75 mm x 50 mm x 1 mm) with the help of a spacer (Corning Cover glass) so that a linear thickness gradient was established (fig. 4a) between the two. The spacer was 180 $\mu$m thick which gave rise to the gradient gel thickness ranging from 0 to 180 $\mu$m over a length of ~ 6.7 cm. This setup was assembled inside a polystyrene petri dish (VWR®, 150mm diameter x 15mm height) where stacks of DI water soaked filter paper were kept on either side of the assembly. As soon as the mixing of the gel solution was complete, it was pipetted out with a sterile transfer pipette and introduced into the wedge formed between the two glass slides. The petri dish was immediately covered by its lid in order to maintain a water vapor rich environment inside.

After allowing the gel to crosslink for two hours, a razor blade was gently inserted in between the upper plate and the spacer till the instability patterns develop all throughout the contact of the gel and the upper plate. These experiments were carried out with the top glass plate coated with either a fluorocarbon silane (FC-10) or a thin (~ 5 nm) polydimethyl siloxane (PDMS) for its easy removal from the gelled film. The patterns were observed using a microscope (Infinity®) equipped with a CCD camera (MTI, CCD-72) and recorded to a computer, which were analyzed later using ImageJ® software. Thin longitudinal strips from the images, obtained at various thickness of the gel film, were taken and the numbers of darker bands cutting across this strip were counted. Wavelength of the instability ($\lambda$) at different thickness ($H$) was obtained by dividing the length of the strip by the number of these bands. We confirmed that this method of measuring $\lambda$ is perfectly consistent with that obtained from the traditional method of Fast Fourier Transforming (FFT) an image provided that the pattern is isotropic and is comparable to the spacing between fingers when a fingering instability is induced by peeling a flexible cantilever from such a hydrogel film (see below). The method used here was quite convenient to analyze the slightly anisotropic spatial patterns resulting from the thickness gradient.

## 3. RESULTS

### 3.1. ESTIMATION OF THE SHEAR MODULUS OF THE GEL

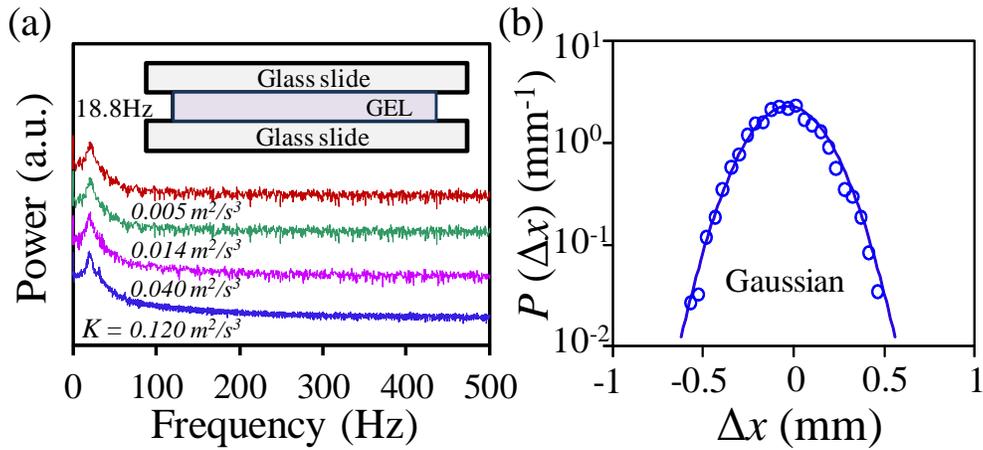

**Figure 1.** (a) The resonance mode of a thin slab of the physically cross-linked hydrogel was obtained by subjecting it to a random excitation parallel to the upper plate while the lower plate was held fixed on the stage of the oscillator. Several power spectra were added and averaged in order to reduce the background noise and improve the peak shape. (b) The probability distribution function of the displacement fluctuation is Gaussian ($K$=0.12 m$^2$/s$^3$) thus emphasizing the linear response of the system. The root mean square (RMS) displacement is 0.4 mm.

The shear elastic modulus of the gel was estimated in two different ways. The first method involved the transverse vibration of a thin slab of the gel of thickness $H$ confined between two flat plates (figure 1a) with a random external noise. The fact that the gel is elastic is evident from the observation that it can support the weight of the upper plate (a glass slide) for an indefinite period of time. From the resonance peak ($\omega$=18.8 Hz) of the gel that behaves like a shear spring, its shear modulus ($\mu$) was estimated using the equation: $2\pi\omega = \sqrt{(\mu A/mH)}$, where $m$ is the mass (4.62 g) of the top vibrating plate, and $A$ is the area of contact between the gel slab and the glass plate. The shear modulus of the gel was found to vary between 42 Pa to 45 Pa. The probability distribution of the shear displacement fluctuation is Gaussian (figure 1b), thus suggesting that the response of the gel is linear, which is reinforced by the fact that the resonance frequency of the gel is independent of the noise strength (figure 1a).

## 3.2. SELF-BRAKING STOKES EXPERIMENT

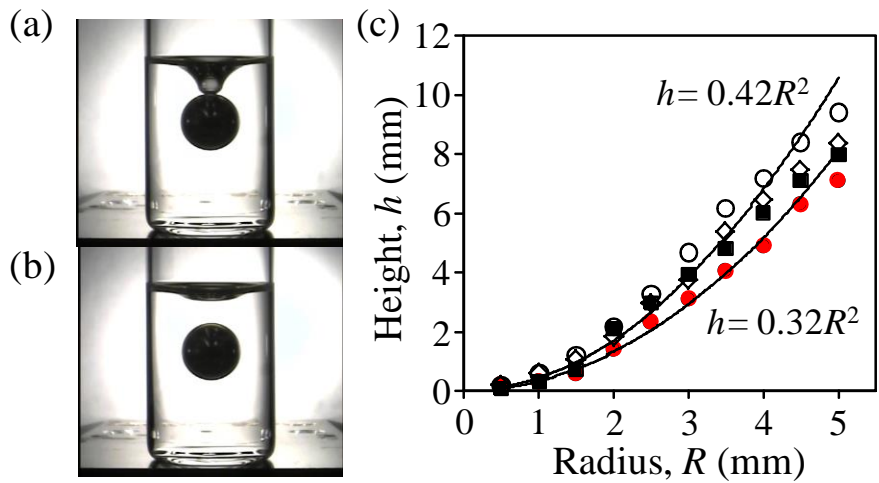

**Figure 2**. (a) A static version of the Stokes experiment, in which a steel ball (R = 5 mm) was released on the surface of a physically cross-linked hydrogel. The ball penetrates deep in the gel and becomes neutrally buoyant, at which stage the elastic shear force on the ball is balanced by the weight of the ball. In (a) the surface of the gel is in contact with the sphere, (b) the gap can be closed by applying couple of drops of an aqueous surfactant solution (1.6 wt% of Brij 35) . (c)

Experiments carried out with balls of different sizes show that the height of submersion increases with the radius of the ball. Although slight differences in the overall behavior is observed with gels prepared on the first day or after 24 hrs, each set of data could be fitted with an equation $h \sim R^2$.

The shear modulus obtained from the vibration experiment discussed as above is adequate for the purpose of analyzing the pattern formation data. The above method of measuring elasticity, however, involves substantial amount of sample preparation time; furthermore, the experiments and analysis are somewhat time consuming. For repeated and routine analysis of the elasticity of the gel, we developed a simple technique in analogy to the classical Stokes experiment, in which a small steel ball is released over the surface of the gel. As shown in figure 2a, the ball submerges itself in the gel by a substantial depth following which the denser steel ball appears to become neutrally buoyant. When the vial is inverted, the ball easily comes out of the gel, which, upon reinsertion, returns to its original position. A larger ball sinks more deeply in the gel than a smaller one because of its greater weight. We also found that the depth to which the ball sinks in the gel is inversely proportional to the degree of gelation, i.e. modulus (Appendix B), the details of which will be published separately. When the experiment is conducted first with the larger ball and then with the smaller one, the latter would reach exactly the same position had the experiment been performed in the reverse order. When an external magnetic field is applied at the bottom of the glass vial, the ball sinks down further only to return to its original position upon the removal of the field. These observations suggested to us that these gels are quite elastic (we thank A. J. Crosby who shared with us some of his observations related to the elastic nature of similar soft hydrogels). When a sufficiently strong magnetic field is applied to the ball, fracture occurs inside the gel, which can then no longer support the weight of the steel ball; the latter simply passes through the gel with a uniform velocity. These results suggest that the gel is most generally not fractured, at least underneath the ball (figure 2a,b), when it stands suspended inside the gel without exhibiting any motion. Since the height of submersion ($h$) decreases with the increase of the shear modulus ($\mu$) of the gel and increases with the radius ($R$) of the sphere, we expect that the shear force due to the accumulated strain would be $C\mu Rh$ in analogy to the Stokes drag in a viscous medium, in which the shear modulus and the height are exchanged with the viscosity and the velocity of the classical Stokes equation, respectively. Using the similarity of the structure of the elastic field (Navier) equation and the Navier-Stokes equation, one can estimate the value of C in the small deformation and linear elastic limit to be $4\pi$ so that the upward force experienced by the spherical ball is $4\pi\mu Rh$ (Appendix C). By balancing this force with that of downward gravity, we obtain an expression for the immersed height as: $h = \Delta\rho g R^2 /(3\mu)$, where $\Delta\rho$ is the excess density of the steel ball surrounded by the hydrogel and $g$ is the gravitational acceleration. Here, we ignore the effect of the wall of the glass vial, which would be important when its inner diameter is comparable to the size of the ball. Figure 2c shows that the depth of submersion ($h$) indeed increases with the radius ($R$) of the sphere in a quadratic fashion. With the data obtained from different measurements, the shear modulus of the gel is estimated to be $58 \pm 3$ Pa at a confidence limit of 95% and by forcing the fitted line to pass through the origin. In order to get a better agreement between this value that (42-45 Pa) estimated from the resonance frequency of the gel (figure 1), the value of C has to close to $6\pi$. We believe that this discrepancy is due to the fact that we used a simplified linear equation of elasticity in the small deformation limit. It would be more appropriate to employ an adequate constitutive relation between stress and strain for such a poro-elastic gel and the coupled effects of the normal stresses in order to account for any putative non-linear effects. Fortunately, at this stage, the discrepancy is not huge and can be taken care of with an empirical correction factor. The value of this simple method is that it allows rapid estimation of the shear modulus of the gel by precluding any dynamics, which can be easily adapted to measure not only the elasticity of a soft gel but also the liquid to solid transition of the gel as gelation is carried out with different amounts of acrylamide.

3.3. SURFACE TENSION OF THE GEL

The need to obtain a direct estimation of the surface tension of the gel is that the material points of its surface can stretch[16] during surface undulation and thus its surface tension may not necessarily be the same as that of liquid water. Direct measurement of the surface tension of a solid is, however, a well-known nuisance in surface physics as it is not usually possible to decouple the surface from the bulk effects. Nevertheless, if the solid is very soft such as the case here and if the perturbation is small, it is possible to estimate its surface tension from the resonance vibration frequency of its free surface. A hemispherical cap of a liquid drop exhibits spherical harmonics with its fundamental frequency scaling with the volume ($V$) as $V^{-0.5}$ provided that the mode is carried out by capillarity[39-45]. For an elasticity driven mode[46], the frequency scales as $V^{-0.33}$.

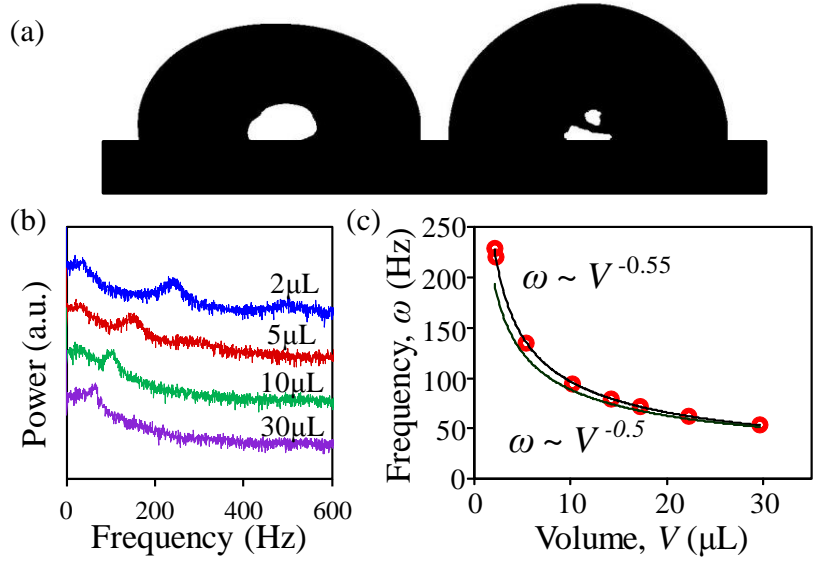

**Figure 3**. (a) The fundamental deformation mode of a *20 μL* hemispherical cap of the hydrogel as obtained from some random frames of the gel vibrating under a random noise. These two frames show the downward (left) and upward (right) deflecting deflections of the surface of the hemispherical gel (b) Power spectra of hemispherical caps of the hydrogel show the resonance modes that depend on the volume of the drop. Several power spectra were added and averaged in order to reduce the background noise and improve the peak shape. (c) The resonance frequency ($\omega$) of the drop (red circles) varies with the volume of the gel following a $V^{-0.55}$ relationship, which is very close to that of water (lower solid line).

Being inspired by such a clear and measurable distinction between the two types of modes, we subjected the hemispherical caps of the physically cross-linked gels to a random vertical vibration[45] and identified the resonance frequency from the power spectrum of its surface fluctuation.

These power spectra (figure 3) show that there is a fundamental vibration mode that varies with the volume of the drop as $\omega \sim V^{-0.55}$, which is similar to that of water in magnitude as well as in character [42-45], i.e. $\omega \sim V^{-0.5}$. The form $\omega \sim V^{-0.55}$ is clearly distinct from the eigen frequencies of the elastic modes[46] for which one expects $\omega \sim V^{-0.33}$. Since the contact angles of the hydrogel caps on the hydrophobic glass supports were 90°, we employ an equation given by Lyubimov et al.[44] according to which the fundamental resonance frequency ($\omega$) of a hemispherical drop is:

$$\omega = \sqrt{\frac{\gamma \tilde{\omega}^2}{6\pi m}} \tag{1}$$

Where $\gamma$ is the surface tension of the drop, $m$ is the mass and $\tilde{\omega}$ is the root of the following equation:

$$\sum_{l=1}^{\infty} \frac{l(4l+1)}{\tilde{\omega}^2 - 4l(2l-1)(l+1)} \left(\frac{(2l-1)!!}{2^l l!}\right)^2 = 0 \tag{2}$$

Using the numerically evaluated value of $\tilde{\omega}$ (4.4268), the fundamental resonance frequency of a hemispherical cap of water of surface tension ~73 mN/m can be estimated from equation 1, which have been rigorously verified in previous experimental studies[43,45]. Figure 3c shows that the experimental resonance frequencies of the gel are quite close to the values predicted from equation 1 for equivalent drops of water. These results encourage us to consider that the surface tension of the gel is very similar to that of pure water, i.e. $\gamma$ (gel) ~ 73 mN/m.

With the above estimates of the elastic modulus and the surface tension of the hydrogel, its elasto-capillary length $\gamma/\mu$ is estimated to be 1.8 mm. Thus, as long as the thickness of the hydrogel film is comparable to or smaller than 1.8 mm, we expect to witness a pronounced elasto-capillarity effect in the interfacial instability. The pattern formation experiments performed with a graded hydrogel bear out this expectation as discussed below.

## 3.4. ELASTO-CAPILLARY INSTABILITY

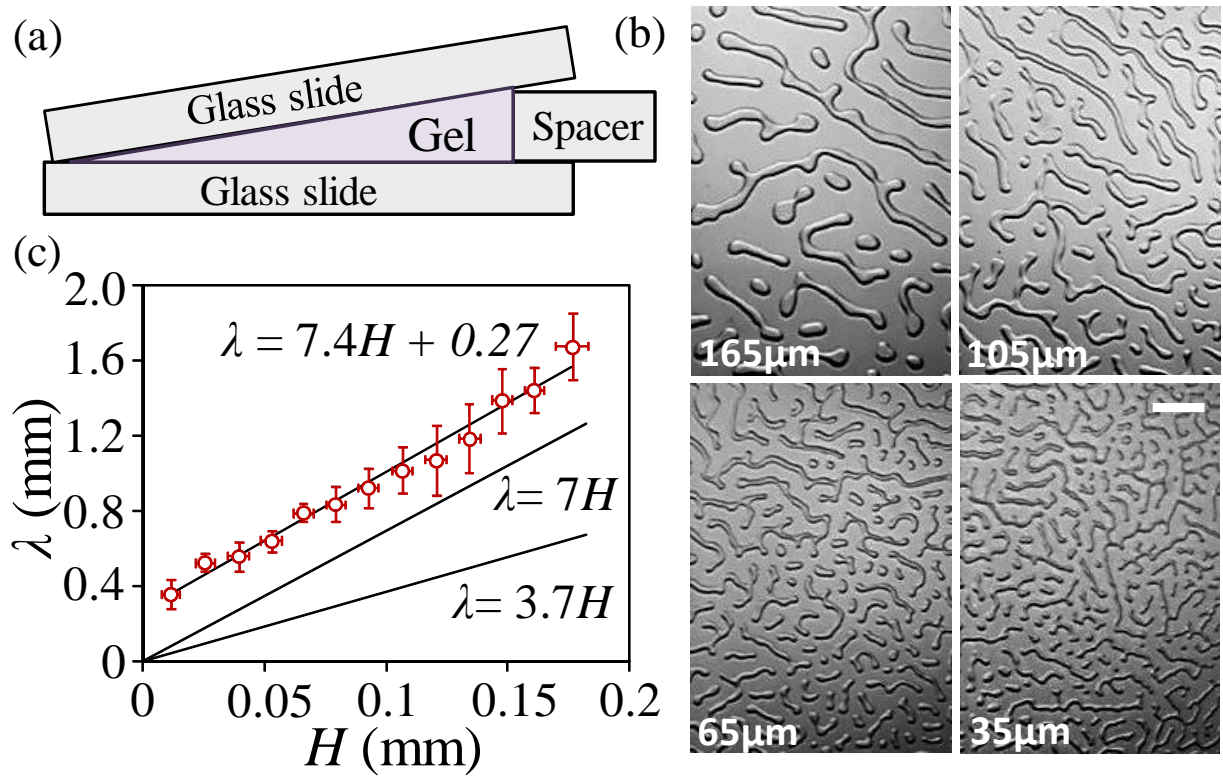

**Figure 4**. (a) A soft hydrogel is confined in the wedge formed by two glass slides. The lower slide is as-received, whereas the upper slide was made hydrophobic by reacting it with a fluorocarbon silane. (b) When the upper plate is slightly lifted from the thicker side of the gel, it detaches partially from the silanized glass thus forming the instability pattern. The white scale bar represents 2 mm. The thickness values are representative of the gel thickness at the center of the micrographs (c) The wavelength (red open circles) of the instability decreases with the thickness of the gel, with its value being much larger than the relationship expected of a purely elastic instability (the solid line, $\lambda=3.7H$).

The adhesion instability experiment could be performed conveniently with a thickness gradient gel that was polymerized inside a wedge shaped geometry (Fig. 4a). The lower slide of the wedge was an untreated glass slide, whereas the upper slide (flexural rigidity of $D = 8$ Nm) was silanized so that it could be easily peeled off the gel from its thicker side by inserting a razor blade underneath the slide resting on the spacer. With a very low wedge angle (0.15°) coupled with a material scale $(D/\mu)^{1/3}$ of deformation (0.6 m) being much larger than any geometric scale of the system, we expect that the entire hydrogel film would be hydrostatically stressed[47] when the upper plate is peeled. This expectation is consistent with the experimental observation that the instability pattern develops spontaneously all throughout the interfacial contact (figures 4b and 5a), with its characteristic wavelength decreasing in proportion to the thickness of the gel. While the gel cured in a wedge geometry simplifies the measurements performed over a significant range of thickness, we emphasize that a gel of uniform thickness also yields similar wavelength of instability as does the gradient gel. For example, while a gel film with an uniform thickness of 0.15 mm yields an instability pattern of $\lambda$ =1.29($\pm$0.06) mm, the section of the gradient gel of similar thickness yields a value $\lambda$ as 1.39($\pm$0.16) mm. This discrepancy is well within the error band of the measurement.

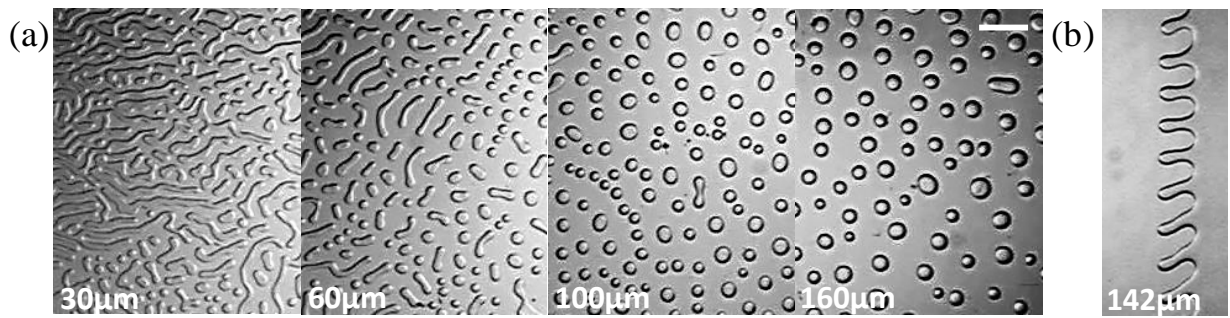

**Figure 5**. (a) This experiment is similar to that of figure 4, except that the upper glass plate was treated with a thin (5 nm) film of polydimethylsiloxane. The white scale bar represents 2 mm. The thickness values are representative of the gel thickness at the center of the micrographs (b) This panel shows the formation of fingering instability, observed at 142$\mu$m thickness, during the peeling of a soft PDMS cantilever (flexural rigidity 2x10$^{-4}$ Nm) from a hydrogel film of varying thickness. The spacing of the fingers is also comparable to the spacing of the bubbles from a film of comparable thickness.

Even though the wavelengths of the instabilities observed with both the FC (fluorocarbon) and the PDMS (polydimethylsiloxane) coated glass slides are comparable, certain differences in their morphologies are evident in figures 4b and 5a. For example, while bubbles are observed with the PDMS coated slide peeling from the thicker part of the gel, interconnected stripes are observed with the FC coated slide. On neither of the two surfaces, the wavelength of the instability follows the standard $\lambda=3.7H$ relationship as was observed previously[21,24-26] with the higher modulus (~ 1 MPa) elastomers. The line joining the data seems to intersect the ordinate axis at a finite value ($0.27 \pm 0.07$ mm), which may tempt one to consider that this intercept is directly related to $\gamma/\mu$. However, for such an interpretation to be valid, the shear modulus of the soft gel has to be about 240 Pa considering that the surface tension of the gel is similar to that of water. This value is considerably greater than that (*ca* 40 Pa) obtained from direct measurements of shear modulus as discussed above. Thus other explanations are sought, in which the effects of the finite compressibility of the film and the elasto-capillarity are explicitly considered.

## 4. DISCUSSIONS

### 4.1. INTERPRETATION OF THE ELASTO-CAPILLARY INSTABILITY

In what follows, we carry out a linear stability analysis of the interface using a relationship between the vertical displacement $w(x)$ of an incompressible elastic layer and the surface normal stress $\sigma(x)$. This relationship between stress and displacement was originally developed by Kerr[48] and later used by Ru[49] to study pattern formation in thin elastic films. Ru[49] showed that such an analysis yields the same result as that of a more formal approach used by Shenoy and Sharma[23] in their studies of the adhesion induced elastic instability. Here we adopt the method of Ru[49], which is easy to use and amenable to the study of the effect of Poisson's ratio in a straightforward manner. While we follow here the lead of Ru[49], there is a technical difference between the method used here and that used by him as well by others[21-23]. In the previous analyses, it was assumed that a long range attractive force (such as a van der Waals force) triggers the instability as the contactor is brought in close vicinity to the soft film much like what was pointed out earlier by Attard and Parker[50]. In our experiment, the gel is already in contact with the substrate and the instability is caused by an external force in the post bonded state. As the long range van der Waals and other interactive forces act in the cohesive zones of detached regions, it is convenient to include the effect of these forces in the work of adhesion ($W_a$). The observed instability is due to the lowering of the potential energy of the system by an external load at the expense of the transverse and longitudinal shear deformations as well as surface undulation.

For a frictionless interface between the gel and the upper substrate, and with a perfect bonding between the film and the lower substrate, Kerr's equation[48,49] takes the following form:

$$w + A_1 H^2 D^2 w + A_2 H^4 D^4 w + A_3 H^6 D^6 w + A_4 H^8 D^8 w = B_1 \frac{H}{E}\sigma + B_2 \frac{H^3}{E} D^2 \sigma + B_3 \frac{H^5}{E} D^4 \sigma + B_4 \frac{H^7}{E} D^6 \sigma \qquad (3)$$

Where $w=w(x)$ is the normal displacement, $E = 2\mu(1+\nu)$ is the Young's modulus, $\nu$ is the Poisson's ratio, $\sigma = \sigma(x)$ is the normal stress, and $D^n$ represents $n^{th}$ derivative with respect to $x$. All the coefficients ($A_i$, $B_i$) are functions of the Poisson ratio ($\nu$) as follows:

$$A_1 = -\frac{1}{1-\nu}, \quad A_2 = \frac{(3-4\nu)}{12(1-\nu)^2}, \quad A_3 = -\frac{(3-4\nu)}{90(1-\nu)^2}, \quad A_4 = \frac{(3-4\nu)}{1260(1-\nu^2)},$$

$$B_1 = \frac{(1+v)(1-2v)}{1-v}, \ B_2 = -\frac{(1+v)(3-4v)}{3(1-v)}, \ B_3 = \frac{(1+v)(3-4v)}{15(1-v)}, \ B_4 = -\frac{2(1+v)(3-4v)}{315(1-v)} \quad (4)$$

Equation 3 can be easily solved for a periodic undulation of the surface ($w = w_o \sin kx$) and the corresponding surface stress as $\sigma = \sigma_o + Mw_o \sin kx$, where $M$ is the stiffness of the film that is determined upon the substitution of perturbed forms of $w$ and $\sigma$ in equation 3 (see also figure 6a). While a closed form relationship[49] can be given for $M$ in terms of $E, H$ and $k$, numerical analysis shows that $MH/E$ follows a power law relationship with $kH$ in the long wave limit (i.e. $kH \leq 1$). For an incompressible film we have an expression for $M$ as follows:

$$M = \frac{E}{H}\left[\frac{1.74}{(kH)^{1.78}}\right] \quad (5)$$

The sum of the surface ($U_S$) and the elastic ($U_E$) energies of the film can now be written down for a 2d deformation as follows:

$$U_S + U_E = \frac{\gamma L_1}{2}\left[\int_0^{L_2}\left(\frac{\partial w}{\partial x}\right)^2 dx\right] + L_1 \int_0^{L_2}\int_0^{w_o}\sigma dw dx \quad (6)$$

where $L_1$ and $L_2$ are the lateral dimensions of the film as shown in figure 6a. Using the periodic perturbations of the surface ($w_o \sin kx$) and stress ($\sigma_o + Mw_o \sin kx$) states, the energy per unit area can be expressed as:

$$\frac{U_S + U_E}{L_1 L_2} = \frac{w_o^2}{4}\left(\gamma k^2 + M\right) \quad (7)$$

In order to carry out the energy analysis of the deformed film, we consider that a constant load is applied on the upper surface so that the system undergoes a net change in the potential energy per unit area as $-\sigma_o w_o$. There is also a decrease of the adhesion energy and a corresponding increase of the elastic and surface energies. Total change of energy per unit area can thus be expressed as

$$\overline{U} = (U_P + U_S + U_E + U_a)/A = -\sigma_o w_o + \frac{w_o^2}{4}(\gamma k^2 + M) - \phi W_a \qquad (8)$$

Where, $\phi$ is the fraction of the surface that is detached. At equilibrium, $\partial \overline{U}/\partial w_o = 0$; we thus obtain an expression for $\overline{U}$ in terms of the applied stress as:

$$\overline{U} = -\frac{\sigma_o^2}{(\gamma k^2 + M)} - \phi W_a \qquad (9)$$

The tendency of the system is to be maximally compliant via interfacial instability, so that the confined film can undergo maximum amount of vertical deflection under a given applied stress. $\overline{U}$ achieves its minimum value when $\gamma k^2 + M$ is minimal with respect to $\lambda$, which leads to an expression for the wavelength of the instability as: $\lambda \approx 4.2 H (\gamma/\mu H)^{0.27}$ for an incompressible film ($\nu=0.5$). This result is consistent with the prediction of our previous scaling analysis ($\lambda \sim H^{3/4}$) and is almost identical to the result [$\lambda \approx 2\pi H (\gamma/3\mu H)^{1/4}$] obtained by Gonuguntla et al[28] in which instability occurs due to attractive forces during the pre-bonding process. This relationship can also be written as $\lambda = 4.2 H / ECa^{0.27}$ where $ECa = \mu H/\gamma$ is the elasto-capillary number that contrasts the expression $\lambda = \pi H / Ca^{0.5}$ applicable for the classical Saffman-Taylor instability[51,52] where $Ca$ is the classical capillary number. The above relationship clearly departs from the conventional $\lambda = 3.7 H$ relationship observed[24-27] previously with less compliant films. While we show below that the experimental results of pattern formation are due to elasto-capillarity, it is tempting to inspect how far this result can be explained by considering a small but finite dilation of the film that favors a long wavelength instability on its own. Calculations with $\gamma=0$ show that the Poisson's ratio has to be in the range of 0.3 so that an instability pattern develops with its wavelength somewhat comparable to experiments. However, the predicted relationship between wavelength and thickness, i.e. $\lambda = 7H$, passes through the origin (0,0) that is markedly different from the experimental observations (fig. 4c).

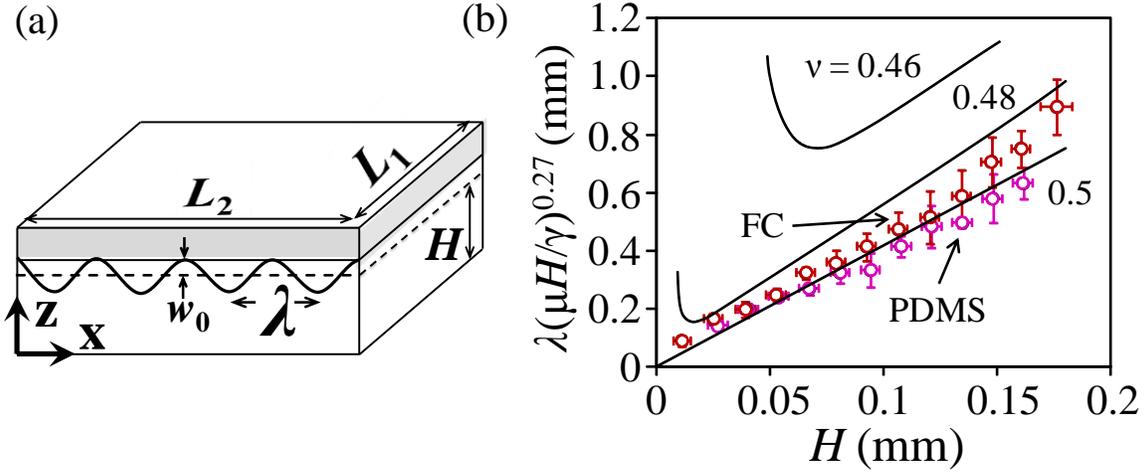

**Figure 6**. (a) Schematics of the geometry used to carry out the linear stability calculation. (b) Calculated values (lines) of the wavelengths are compared with experiments (filled symbols). The red circles represent the data obtained with the peeling of a fluorocarbon silane treated glass from the PAM hydrogel, whereas the pink circles represent the data obtained with a PDMS coated glass peeling from a PAM hydrogel. All the wavelengths are re-scaled by multiplying it with $(\mu H/\gamma)^{0.27}$ using the value of $\gamma$ as 0.073 mN/m and $\mu$ as 40 Pa. The calculations are performed with three different Poisson's rations (0.5, 0.48 and 0.46) as indicated in the inset of the figure.

Now, considering that the hydrogel to be incompressible, the values of the wavelength as obtained by minimizing $\gamma k^2 + M$ are plotted in figure 6b as a function of $H$ after rescaling the wavelength as $\lambda(\mu H/\gamma)^{0.27}$. The data obtained with both the FC and PDMS coated glass slides cluster around this theoretical line corresponding to $\nu = 0.5$ with their intercepts on the ordinate axes being $0.016 \pm 0.044$ and $0.041 \pm 0.032$ respectively at a confidence limit of 95%. The significant error bands associated with the estimates of these intercepts prevent us from making definitive comments about the Poisson's dilation. Calculations also show that if the putative

Poisson effect were present, it would have strongly influenced the wavelength of the interfacial patterns in the very thin film region. While the scaling $\lambda(\mu H/\gamma)^{0.27} \sim H$ would still be obeyed by the thicker films with a finite dilatation, the thinner films would exhibit such long wavelengths that the interface may not roughen at all below a critical thickness. The thickness of the film where this transition occurs is predicted to be inversely proportional to the Poisson's ratio, i.e. while this transition occurs at about $H = 10$ μm for $\nu = 0.48$, it occurs at about $H = 45$ μm for $\nu = 0.46$. Since the experimental instabilities are observed with $H$ as low as 10 μm, we feel that the experimental data should be compared with the theoretical analysis performed with the Poisson's ratio being very close to 0.5.

We summarize by stating that we found convincing evidence of elasto-capillarity in adhesion induced pattern formation between a solid contactor and an ultra-soft gel film. The rescaled wavelength of instability varies fairly linearly with the film thickness, which is in good agreement with the theoretical analysis. Some differences in the morphological patterns of the instability are observed with the FC and the PDMS coated surfaces, which may arise due to some well-known differences[53] in hysteresis, adhesion, and friction of these surfaces. In other words, the slip boundary condition that is intrinsic in equation 3 may be different for the FC and PDMS surfaces.

4.2. ELASTO-CAPILLARITY IN ADHESIVE FRACTURE

Understanding the nature of elasto-capillary instability is also important in estimating the adhesive fracture stress ($\sigma^*$) of a flat ended rigid indenter from a thin confined film of thickness $H$. The subject has its origin in the classic work of Kendall[54], who laid down the foundation for such an analysis by proposing that the adhesive fracture stress of a very thin film undergoing volume dilation is $\sigma^* \sim (K_b W_a / H)^{0.5}$, where $K_b$ is the bulk modulus and $W_a$ is the work of adhesion. Later it was argued in the literature[25-27,55,56] that a thin film can bypass the above mode of interfacial separation in different ways, one being the elastic instability[25-27]. In such a case, Kendall's equation assumes a form: $\sigma^* \sim (\mu W_a / H)^{0.5}$ where the bulk modulus $K_b$ is replaced by the shear modulus $\mu$. This would be the case for a large elastocapillary number, i.e. $\mu H \gg \gamma$. However, when $\mu H$ is comparable to or less than $\gamma$, certain non-trivial regimes may appear as follows.

In order to develop the premise for this discussion, we express $k$ and $M$ in terms of the wavelength of the instability $\lambda$ so as to obtain the scaled total energy ($U = \bar{U}A$) of the system (eq. 9) as follows:

$$U \sim -\frac{F^2 \lambda^2}{\gamma A} - W_a A \tag{10}$$

Where, $F$ is the applied force, which is kept constant during the fracture process. With an expression for $\lambda$ as $\lambda \sim (\gamma H^3 / \mu)^{1/4}$, the pull-off stress is given by the instability conditions: $\partial U / \partial A)_F = 0$ and $\partial^2 U / \partial A^2 < 0$, which lead to eq. 11.

$$\sigma^* \sim \left(\frac{\gamma W_a^2 \mu}{H^3}\right)^{\frac{1}{4}} \quad \text{or} \quad \sigma^* \sim \left(\frac{\gamma}{\mu H}\right)^{0.25} \left(\frac{W_a \mu}{H}\right)^{0.5} \tag{11}$$

We thus have a situation where the adhesive stress depends more strongly on the thickness of a film than the usual case [$\sigma^* \sim (W_a \mu / H)^{0.5}$] of an elastic instability driven crack formation and the subsequent rupture of the contact. A non-trivial case may manifest with a very thin film, in which the surface tension no longer allows auto-roughening of the surface of a dilatable film (i.e. $\nu < 0.5$). The critical stress to fracture could then depend on the bulk modulus ($K_b$) as $\sigma^* \sim (W_a K_b / H)^{0.5}$, which is the classic equation proposed by Kendall[54] over forty years ago. Many practical soft adhesives however are viscoelastic[57] and thus additional improvisations would be needed. Further research needs to be carried out to verify these predictions with ultra-soft and/or ultra-thin elastic adhesives sandwiched between rigid places, which may be important in the construction of an appropriate phase diagram for thin film adhesion. These results may also be relevant in tribological settings involving soft materials for which there are suggestions[58] and ample evidences[59] of the roles played by the interfacial instabilities.

## 5. CONCLUSIONS

The conclusions of this work are as follows:

1. Direct measurements of the surface tension and the elastic modulus led to the prediction of a rather large and macroscopically realizable elasto-capillary length in a soft elastic hydrogel.

2. Experimental and theoretical analysis corroborate that the long wave instability observed in a soft elastic film is the consequence of a large elasto-capillary length with some non-trivial effects arising from the Poisson's dilation.

3. It is further proposed that a long wavelength instability can affect the mode of fracture as well as the adhesive strength of a soft adhesive confined between two rigid substrates in non-trivial ways.

# APPENDIX

## A. SCALING ANALYSIS OF ELASTO-CAPILLARY INSTABILITY

Here we show a scaling analysis[25,26] to obtain an expression for the spatial wavelength of the elasto-capillary instability. We begin by writing the sum of the elastic and surface energies (per unit area basis) in terms of the horizontal ($u$) and the vertical ($w$) displacements as follows:

$$\overline{U} \sim \mu H \left[ \left(\frac{\partial u}{\partial z}\right) + \left(\frac{\partial w}{\partial x}\right) \right]^2 + \gamma \left(\frac{\partial w}{\partial x}\right)^2, \tag{12}$$

We wish to reduce equation 12 at the scaling level by choosing the characteristic length scales in the horizontal and vertical directions as the spatial wavelength ($\lambda$) and the thickness of the film ($H$) respectively. Now, taking the amplitude of the perturbation as $w_o$, we have $\partial u / \partial z \sim u / H$ and $\partial w / \partial x \sim w_o / \lambda$. The maximum horizontal displacement scales along the $x$ direction can be obtained from the equation of continuity ($\partial u / \partial x + \partial w / \partial z = 0$) or $u / \lambda \sim w_o / H$, which leads to $u \sim w_o \lambda / H$. Equation 12 can now be written at the scaling level as $\overline{U} \sim \mu H w_o^2 (\lambda / H^2 + 1/\lambda)^2 + \gamma w_0^2 / \lambda^2$. Minimization of $\overline{U}$ with respect to $\lambda$ yields the desired relation: $\lambda \sim H (1 + \gamma / \mu H)^{1/4}$.

*Note:* MKC is indebted to L. Mahadevan, who prompted him to derive the above equation using a scaling analysis *ca* 2002.

## B. A STEEL BALL STANDS STILL INSIDE A SOFT HYDROGEL

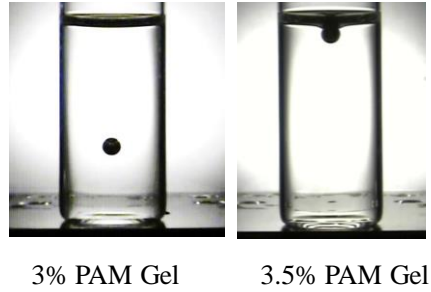

3% PAM Gel    3.5% PAM Gel

**Figure 7.** The submerged height of a steel ball (4 mm diameter) inside a polyacrylamide (PAM) gel is inversely proportional to the concentration of PAM used for the gelation reaction.

## C. ELASTIC STOKES EQUATION

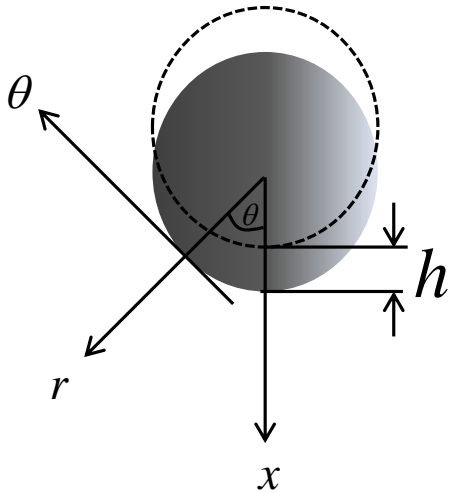

**Figure 8.** Schematic of a rigid sphere undergoing a small displacement ($h$) in an elastic medium.

Conventional notations of stress, strain and displacements will be used in the following derivation of the elastic counterpart of the Stokes equation. The displacement field that satisfies the stress equilibrium relations are[60]:

$$u_r = h\cos\theta \frac{R}{r}$$

$$u_\theta = -h\sin\theta \frac{R}{2r}$$

(13)

The normal components of the strain and stress are given in equations 14 and 15 respectively:

$$u_{rr} = \frac{\partial u_r}{\partial r} = -h\cos\theta \frac{R}{r^2}$$
$$u_{\theta\theta} = \frac{1}{r}\frac{\partial u_\theta}{\partial \theta} + \frac{u_r}{r} = \frac{1}{r}\left(-h\cos\theta\frac{R}{2r} + h\cos\theta\frac{R}{r}\right) = h\cos\theta\frac{R}{2r^2}$$
$$u_{\phi\phi} = \frac{u_\theta}{r}\cot\theta + \frac{u_r}{r} = \frac{1}{r}\left(-h\cos\theta\frac{R}{2r} + h\cos\theta\frac{R}{r}\right) = h\cos\theta\frac{R}{2r^2}$$

(14)

$$\sigma_{rr} = -p + 2\mu u_{rr} = -p + 2\mu\left(-h\cos\theta\frac{R}{r^2}\right)$$
$$\sigma_{\theta\theta} = -p + 2\mu u_{\theta\theta} = -p + 2\mu\left(h\cos\theta\frac{R}{2r^2}\right)$$
$$\sigma_{\phi\phi} = -p + 2\mu u_{\phi\phi} = -p + 2\mu\left(h\cos\theta\frac{R}{2r^2}\right)$$

(15)

Now setting the stresses tangential to the surface of the ball to zero, i.e. $\sigma_{\theta\theta} = \sigma_{\phi\phi} = 0$, we obtain an expression for the pressure (*p*) as follows:

$$p = \mu h \cos\theta \frac{R}{r^2}$$

(16)

The normal stress acting on the sphere then is:

$$-\sigma_{rr} = 3\mu h \cos\theta \frac{R}{r^2}$$

(17)

Thus, the total normal load acting on the sphere is:

$$F = \int_0^\pi \left(-\sigma_{rr}\big|_{r=R}\cos\theta\right)(Rd\theta)(2\pi R \sin\theta)$$

(18)

Substituting equation 17 in equation 18, we have

$$F = 6\pi\mu hR \int_0^\pi \cos^2\theta \sin\theta \, d\theta = 4\pi\mu hR$$

(19)

We recently learned that Cantat and Pitois[61] used a similar method to estimate the shear modulus of foam, in which a sphere of known radius was inserted in a foam and the resulting force was measured as a function of the insertion depth. They presented the same equation 20 in their paper[60]. The derivation given here was provided by Animangsu Ghatak. This model, however, is in small deformation limit. More elaborate derivation using a non-linear field theory will be presented in future.

ACKNOWLEDGEMENTS

We thank Animangsu Ghatak (Indian Institute of Technology at Kanpur) for many discussions. Ghatak, a former PhD student in our group, pioneered the studies of adhesion induced instabilities.